\documentclass[11pt]{amsart}
\usepackage{latexsym,amssymb,amsmath,amscd,amsthm}\usepackage{amsfonts}
\numberwithin{equation}{section}
\theoremstyle{remark}

\newcommand{\bq}{\begin{equation}}
\newcommand{\bea}{\begin{array}}
\newcommand{\eea}{\end{array}}

\newcommand{\ga}{\alpha}
\newcommand{\gep}{\epsilon}
\newcommand{\gD}{\Delta}

\newcommand{\gb}{\beta}

\newcommand{\mc}{\mathcal}

\newcommand{\gG}{\Gamma}

\newcommand{\gz}{\zeta}

\newcommand{\pp}{\partial}

\newcommand{\na}{\nabla}

\newcommand{\bs}{\blacksquare}

\newcommand{{\DDD}}{D\!\!\!\!\!\!-}
\title{ON A FRACTIONAL QUANTUM POTENTIAL}

\author{Robert Carroll\\University of Illinois, Urbana, IL 61801}

\date{March, 2012\thanks{email: rcarroll@math.uiuc.edu}}

\begin{document}

%\tableofcontents

\bibliographystyle{plain}

%\begin{document}

\maketitle

%\tableofcontents

%\tableofcontents

\section{INTRODUCTION}
\renewcommand{\theequation}{1.\arabic{equation}}
\setcounter{equation}{0}

For fractals we refer to \cite{falc,mndl} and for differential equations cf.
also \cite{kiga,lask,strz,zak,zaak}.  The theme of scale relativity as in \cite{aunl,clnl,cenl,nott,notl,noce,nllr}
provides a profound development of differential calculus involving fractals (cf. also
the work of Agop et al in the journal Chaos, Solitons, and Fractals) and for interaction with
fractional calculus we mention \cite{ccgn,hrmn,nbtr,trsv,trzy}.  There are also connections with the
Riemann zeta function which we do not discuss here (see e.g. \cite{mnkw}).  Now the recent
paper \cite{kblv} of Kobelev describes a Leibnitz type fractional derivative and one can relate
fractional calculus with fractal structures as in \cite{ccgn,klgl,rowt,trsv,trzy} for example.  On the
other hand scale relativity with Hausdorff dimension 2 is intimately related to the Schr\"odinger 
equation (SE) and quantum mechanics (QM) (cf. \cite{notl}).  We show now that if one
can write a meaningful Schr\"odinger equation with Kobelev derivatives ($\ga$-derivatives) then there will be a 
corresponding fractional quantum potential (QP) (see e.g. \cite{hrmn,lask,trsv,trzy} for a related fractional equation and recall
that the classical wave function for the SE has the form $\psi=Rexp(iS/\hbar)$).
\\[3mm]\indent
Going now to \cite{kblv} we recall the Riemann-Liouville (RL) type fractional operator
(assumed to exist here)
\bq\label{1.1}
{}_cD_z^{\ga}[f(z)]=\left\{
\begin{array}{cc}
\frac{1}{\gG(-\ga)}\int _c^z(z-\gz)^{-\ga-1}f(\gz)d\gz & c\in{\bf R},\,\,Re(\ga)<0\\
\frac{d^m}{dz^m}{}_cD_z^{\ga-m}[f(z)] & m-1\leq \Re{\ga}<m
\end{array}\right.
\end{equation}
(the latter for $m\in{\bf N}=\{1,2,3,\cdots\}$).  For $c=0$ one writes $({\bf 1A})\,\,{}_0D^{\ga}_z[f(z)]=D_z^{\ga}[f(z)]$ as in the classical RL operator of order $\ga$ (or $-\ga$).  Moreover when $c\to\infty$ (1.1) may be identified with the familiar Weyl fractional derivative (or integral) of order $\ga$ (or $-\ga$).
An ordinary derivative corresponds to $\ga=1$ with $({\bf 1B})\,\,(d/dz)[f(z)]=D_z^{\ga}[f(z)]$.  The
binomial Leibnitz rule for derivatives is
\bq\label{1.2}
D_z^1[f(z)g(z)]=g(z)D_z^1[f(z)]+f(z)D_z^1[g(z)]
\end{equation}
whose extension in terms of RL operators $D_z^{\ga}$ has the form
\bq\label{1.3}
D_z^{\ga}[f(z)g(z)]=\sum_{n=0}^{\infty}
\left(\begin{array}{c}
\ga\\
n
\end{array}\right)
D_z^{\ga-n}[f(z)]D_z^n[g(z)]
\end{equation}
$$\left(\begin{array}{c}
\ga\\
k
\end{array}\right)=\frac{\gG(\ga+1)}{\gG(\ga-k+1)\gG(k+1)};\,\,\,\ga,\,k\in{\bf C}$$
The infinite sum in (1.3) complicates things and the binomial Leibnitz rule of \cite{kblv} will
simplify things enormously.  Thus consider first a momomial $z^{\gb}$ so that
\bq\label{1.4}
D_z^{\ga}[z^{\gb}]=\frac{\gG(\gb+1)}{\gG(\gb-\ga+1)}z^{\gb-\ga};\,\,\Re(\ga)<0;\,\,\Re(\gb)>-1
\end{equation}
Thus the RL derivative of $z^{\gb}$ is the product
\bq\label{1.5}
D^{\ga}_z[z^{\gb}]=C^*(\gb,\ga)z^{\gb-\ga};\,\,C^*(\gb,ga)=\frac{\gG(\gb+1)}{\gG(\gb-\ga+)}
\end{equation}

\indent
Now one considers a new definition of a fractional derivative referred to as an $\ga$
derivative in the form
\bq\label{1.6}
\frac{d_{\ga}}{dz}[z^{\gb}]=d_{\ga}[z^{\gb}]=C(\gb,\ga)z^{\gb-\ga}
\end{equation}
This is required to satisfy the Leibnitz rule (1.2) by definition, given suitable conditions on 
$C(\gb,\ga)$.  Thus first $({\bf 1C})\,\,z^{\gb}=f(z)g(z)$ with $f(z)=z^{\gb-\gep}$ and
$g(z)=z^{\gep}$ for arbitrary $\gep$ the application of (1.3) implies that
\bq\label{1.7}
\frac{d_{\ga}}{dz}[z^{\gb}]=z^{\gep}\frac{d_{\ga}}{dz}z^{\gb-\gep}+z^{\gb-\gep}\frac
{d_{\ga}}{dz}z^{\gep}=z^{\gep}C(\gb-\gep,\ga)z^{\gb-\gep-\ga}+
\end{equation}
$$+z^{\gb-\gep}C(\gep,\ga)z^{\gep-\ga}=[C(\gb-\gep,\ga)+C(\gep,\ga)]z^{\gb-\ga}$$
Comparison of (1.6) and (1.7) yields $({\bf 1D})\,\,C(\gb-\gep,\ga)+C(\gep,\ga)=C(\gb,\ga)$.
To guarantee (1.2) this must be satisfied for any $\gb,\,\gep,\,\ga$.  Thus ({\bf 1D}) is the basic functional
equation and its solution is $({\bf 1E})\,\,C(\gb,\ga)=A(\ga)\gb$.  Thus for the validity of the Leibnitz
rule the $\ga$-derivative must be of the form
\bq\label{1.8}
d_{\ga}[z^{\gb}]=\frac{d_{\ga}}{dz}[z^{\gb}]=A(\ga)\gb z^{\gb-\ga}
\end{equation}
One notes that $C^*(\gb,\ga)$ in (1.5) is not of the form ({\bf 1E}) and the RL operator $D_z^{\ga}$ does
not in general possess a Leibnitz rule.  One can assume now that $A(\ga)$ is arbitrary and $A(\ga)=1$ is chosen.  Consequently for any $\gb$
\bq\label{1.9}
\frac{d_{\ga}}{dz}z^{\gb}=\gb z^{\gb-\ga};\,\,\frac{d_{\ga}}{dz}z^{\ga}=\ga;\,\,\frac{d_{\ga}}{dz}
z^0=0
\end{equation}

\indent
Now let K denote an algebraically closed field of characteristic 0 with $K[x]$ the corresponding 
polynomial ring and $K(x)$ the field of rational functions.  Let $F(z)$ have a
Laurent series expansion about 0 of the form
\bq\label{1.10}
F(z)=\sum_{-\infty}^{\infty}c_kz^k;\,\,F_{+}(z)=\sum_0^{\infty}c_kz^k;\,\,F_{-}(z)=\sum_{-\infty}^{-1}
c_kz^k;\,\,c_k\in K
\end{equation}
and generally there is a $k_0$ such that $c_k=0$ for $k\leq k_0$.
The standard ideas of differentiation hold for $F(z)$ and formal power series form a ring $K[[x]]$
with quotient field $K((x))$ (formal Laurent series).  One considers now the union $({\bf 1F})\,\,
K<<x>>=\cup_1^{\infty}K((x^{1/k}))$.  This becomes a field if we set 
\bq\label{1.11}
x^{1/1}=x,\,\,x^{m/n}=(x^{1/n})^m
\end{equation}
Then $K<<x>>$ is called the field of fractional power series or the field of Puiseux series.  If 
$f\in K<<x>>$ has the form $({\bf 1G})\,\,f=\sum_{k_o}^{\infty}c_kx^{m_k/n_k}$ where $c_1\ne 0$ and
$m_k,\,n_k\in {\bf N}=\{1,2,3,\cdots\},\,\,(m_i/n_i)<(m_j/n_j)$ for $i<j$ then the order is $({\bf 1H})\,\,
O(f)=m/n$ where $m=m_1,\,\,n=n_1$ and $f(x)=F(x^{1/n})$.  Now given n and $z$ complex
we look at functions
\bq\label{1.12}
f(z)=\sum_{-\infty}^{\infty}c_k(z-z_0)^{k/n}=f_{+}(z)+f_{-}(z);\,\,f_{+}(z)=\sum_0^{\infty}
c_k(z-z_0)^{k/n},
\end{equation}
$$f_{-}(z)=\sum_{-\infty}^{-1}c_k(z-z_0)^{k/n};\,\,c_k=0\,\,(k\leq k_0)$$
(cf. \cite{kblv} for more algebraic information - there are some misprints).
\\[3mm]\indent
One considers next the $\ga$-derivative for a basis $({\bf 1I})\,\,\ga=m/n;\,\,0<m<n;\,\,m,n\in {\bf N}=\{1,2,3,\cdots\}$.  The $\ga$-derivative of a Puiseux function of order $O(f)=1/n$ is again
a Puiseux function of order $(1-m)/n$.  For $\ga=1/n$ we have
\bq\label{1.13}
f_{+}=\sum_0^{\infty}c_kz^{k/n}=\sum_0^{\infty}c_kz^{\gb};\,\,\gb=\gb(k)=\frac{k}{n}
\end{equation}
leading to
\bq\label{1.14}
\frac{d_{\ga}}{dz}f_{+}(z)=\sum_1^{\infty}\ga \gb c_kz^{(k-1)/n}=\sum_0^{\infty}c_{p+1}\ga\gb  z^{p/m};
\end{equation}
$$\frac{d_{\ga}}{dz}f_{-}(z)=\sum_{-\infty}^{-1}c_k\ga\gb z^{(k-1)/n}=\sum_{-\infty}^{-2}c_{p+1}\ga\gb z^{p/n}=\sum_{-\infty}^{-1}\hat{c}_pz^{p/n};\,\,\hat{c}_{-1}=0$$
Similar calculations hold for $\ga=m/n$ (there are numerous typos and errors in indexing in \cite{kblv}
which we don't mention further).  The crucial property however is the Leibnitz rule
\bq\label{1.15}
\frac{d_{\ga}}{dz}(fg)=g\frac{d_{\ga}}{dz}f +f\frac{d_{\ga}}{dz}g;\,\,(d_{\ga}\sim\frac{d_{\ga}}{dz})
\end{equation}
which is proved via arguments with Puiseux functions.  This leads to the important chain rule
\bq\label{1.16}
\frac{d_{\ga}}{dz}F(g_i(z))=\sum\frac{\pp F}{\pp g_k}\frac{d_{\ga}}{dz}g_k(z)
\end{equation}
Further calculation yields (again via use of Puiseux functions)
\bq\label{1.17}
\frac{d^m_{\ga}}{dz^m}\left[\frac{d^{\ell}_{\ga}}{dz^{\ell}}f\right]=\frac{d^{\ell}_{\ga}}{dz^{\ell}}
\left[\frac{d^m_{\ga}}{dz^m}f\right]
\end{equation}
\bq\label{1.18}
\int f(z)d_{\ga}z=\sum_0^{\infty}\int z^{\gb}d_{\ga}z;\,\,\int z^{\gb}d_{\ga}z=\frac{z^{\gb+\ga}}{\gb+\ga}
\end{equation}
\bq\label{1.19}
\frac{d_{\ga}}{dz}\int f(z)d_{\ga}z=f(z)=\int \frac{d_{\ga}}{dz}d_{\ga}z
\end{equation}
where $d_{\ga}z$ here is an integration symbol here).
\\[3mm]\indent
The $\ga$-exponent is defined as 
\bq\label{1.20}
E_{\ga}(z)=\sum_0^{\infty}\frac{(z^{\ga}/\ga)^k}{\gG(\ga+1)}=
\end{equation}
$$=exp\left(\frac{z^{\ga}}{\ga}\right);\,\,E_1(z)=e^z;\,\,
E_{\ga}(0)=1\,\,(0<\ga,1)$$
The definition is motivated by the fact that $E_{\ga}(z)$ satisfies the $\ga$-differential equation
$({\bf 1J})\,\,(d_{\ga}/dz)E_{\ga}(z)=E_{\ga}(z)$ with $E_{\ga}(0)=1$.  This is proved by term to
term differentiation of (1.20).  It is worth mentioning that $E_{\ga}(z)$ does not possess the
semigroup property $({\bf 1K})\,\,E_{\ga}(z_1+z_2)\ne E_{\ga}(z_1)E_{\ga}(z_2)$.

\section{FRACTALS AND FRACTIONAL CALCULUS}
\renewcommand{\theequation}{2.\arabic{equation}}
\setcounter{equation}{0}

For relations between fractals and fractional calculus we refer to \cite{ccgn,eynk,klgl,rlwq,rowt,trsv,trzy,yxjn}.  In \cite{ccgn} for example one assumes time
and space scale isotropically and writes $[x^{\mu}]=-1$ for $\mu=0,1,\cdots,D-1$ and the
standard measure is replaced by $({\bf 2A})\,\,d^Dx\to d\rho(x)$ with $[\rho]=-D\ga\ne -D$
(note $[\,\,]$ denotes the engineering dimension in momentum units).  Here $0<\ga<1$ is a
parameter related to the operational definition of Hausdorff dimension which determines the
scaling of a Euclidean volume (or mass distribution) of characteristic size R (i.e. $V(R)\propto
R^{d_H}$).  Taking $\rho\propto d(r^{D\ga})$ one has $({\bf 2B})\,\,V(R)\propto \int d\rho_{Euclid}(r)=
\propto \int_0^Rdr r^{D\ga -1}\propto R^{D\ga}$, showing that $\ga=d_H/D$.  In general as 
cited in \cite{ccgn} the Hausdorff dimension of a random process (Brownian motin) described by a fractional differintegral is proportional to the order $\ga$ of the differintegral.  The same relation holds for deterministic fractals and in general the fractional differintegration of a curve changes its Hausdorff
dimension as $d_H\to d_H+\ga$.  Moreover integrals on "net fractals" can be approximated by the left sided RL fractional of a function $L(t($ via
\bq\label{2.1}
\int_0^{\bar{t}}d\rho(t)L(t)\propto {}_0I^{\ga}_{\bar{t}}L(t)=\frac{1}{\gG(t)}\int_0^{\bar{t}}dt
(\bar{t}-t)^{\ga-1}L(t);
\,\,\rho(t)=\frac{\bar{t}^{\ga}-(\bar{t}-t)^{\ga}}{\gG(\ga+1)}
\end{equation}
where $\ga$ is related to the Hausdorff dimension of the set (cf. \cite{rlwq}.  Note that a change
of variables $t\to \bar{t}-t$ transforms (2.1) to
\bq\label{2.2}
\frac{1}{\gG(\ga)}\int_0^tdt t^{\ga-1}L(\bar{t}-t)
\end{equation}
The RL integral above can be mapped into a Weyl integral for $\bar{t}\to\infty$.  Assuming
$lim_{\bar{t}\to\infty}$ the limit is formal if the Lagrangian $L$ is not autonomous and one assumes therefore that $lim_{\bar{t}\to\infty}L(\bar{t}-t)=L[q(t),\dot{q}(t)]$ (leading to a Stieltjes field theory
action).  After constructing a ``fractional phase space" this analogy confirms the interpretation of the order of the fractional integral as the Hausdorff dimension of the underlying fractal (cf. \cite{trsv}).
\\[3mm]\indent
Now for the SE we go to \cite{hrmn,lask,trsv,trzy}.  Thus from \cite{lask} (1009.5533) one looks at a
Hamiltonian operator 
\bq\label{2.3}
H_{\ga}(p,r)=D_{\ga}|p|^{\ga}+V(r)\,\,(1<\ga\leq 2)
\end{equation}
When $\ga=2$ one has $D_2=1/2m$ which gives the standard Hamiltonian operator $({\bf 2C})\,\,
\hat{H}(\hat{p},\hat{r})=(1/2m)\hat{p}^2+\hat{V}(\hat{r]}$.  Thus the fractional QM (FQM) based on the Levy path integral generalizes the standard QM based on the Feynman integral for example.  This means that
the path integral based on Levy trajectories leads to the fractional SE.  For Levy index $\ga=2$ the Levy
motion becomes Brownian motion so that FQM is well founded.  Then via (2.2) one obtains a fractional SE (GSE) in the form
\bq\label{2.4}
i\hbar\pp_t\psi=D_{\ga}(-\hbar^2\gD)^{\ga/2}\psi+V(r)\psi\,\,\,(1<\ga\leq 2)
\end{equation}
with 3D generalization of the fractional quantum Riesz derivative $(-\hbar^2\gD)^{\ga/2}$ introduced via
\bq\label{2.5}
(-\hbar^2\gD)^{\ga/2}\psi(r,t)=\frac{1}{(2\pi\hbar)^3}\int d^3p e^{\frac{ipr}{\hbar}}|p|^{\ga}\phi(p,t)
\end{equation}
where $\phi$ and $\psi$ are Fourier transforms.  The 1D FSE has the form
\bq\label{2.6}
i\hbar\pp_t\psi(x,t)=-D_{\ga}(\hbar\na)^{\ga}\psi+V\psi\,\,\,(1<\ga\leq 2)
\end{equation}
The quantum Riesz fractional derivative is defined via
\bq\label{2.7}
(\hbar\na)^{\ga}\psi(x,t)=-\frac{1}{2pi\hbar}\int_{-\infty}^{\infty}dp\,e^{\frac{ipx}{\hbar}}|p|^{\ga}\phi(p,t)
\end{equation}
where
\bq\label{2.8}
\phi(p,t)=\int_{-\infty}^{\infty}dx\,e^{\frac{-ixt}{\hbar}}\psi(x,t)
\end{equation}
with the standard inverse.  Evidently (2.6) can be written in operator form as $({\bf 2D})\,\,i\hbar\pp_t\psi=H_{\ga}\psi;\,\,H_{\ga}=-D_{\ga}(\hbar\na)^{\ga}+V(x)$
\\[3mm]\indent
In \cite{hrmn} (0510099) a different approach is used involving the Caputo derivatives (where ${}_c^{+}
D(x)k=0$ for $k= constant$.  Here for $({\bf 2E})\,\,f(kx)=\sum_0^{\infty}a_n(kx)^{n\ga}$ one writes
($D\to\bar{D}$)
\bq\label{2.9}
{}_c^{+}f(kx)=k^{\ga}\sum_0^{\infty}a_{n+1}\frac{\gG(1+(n+1)\ga)}{\gG(1+n\ga)}(kx)^{n\ga}
\end{equation}
Next to extend the definition to negative reals one writes 
\bq\label{2.10}
x\to\bar{\chi}(x)=sgn(x)|x|^{\ga};\,\,\bar{D}(x)=sgn(x){}_c^{+}D(|x|)
\end{equation}
There is a parity tranformation $\Pi$ satisfying $({\bf 2F})\,\,\Pi\bar{\chi}(x)=-\bar{\chi}(x)$ and 
$\Pi\bar{D}(x)=-\bar{D}(x)$.  Then one defines $({\bf 2G})\,\,f(\bar{\chi}(kx))=\sum_0^{\infty}
a_n\bar{\chi}^n(kx)$ with a well defined derivative
\bq\label{2.11}
\bar{D}f(\bar{\chi}(kx))=sgn(k)|k|^{\ga}\sum_0^{\infty}a_{n+1}
\frac{\gG(1+(n+1)\ga)}{\gG(1+n\ga)}\bar{\chi}^n(kx)
\end{equation}
This leads to a Hamiltonian $H^{\ga}$ with
\bq\label{2.12}
H^{\ga}=-\frac{1}{2}mc^2\left(\frac{\hbar}{mc}\right)^{2\ga}\bar{D}^i\bar{D}_i+V(\hat{X}^1,
\cdots,\hat{X}^i,\cdots,\hat{X}^{3N})
\end{equation}
with a time dependent SE
\bq\label{2.13}
H^{\ga}\Psi=\left[-\frac{1}{2}mc^2\left(\frac{\hbar}{mc}\right)^{2\ga}\bar{D}^i\bar{D}_i
+V(\hat{X}^1,\cdots,\hat{X}^i,\cdots\hat{X}^{3N})\right]\Psi=i\hbar\pp_t\Psi
\end{equation}

\section{THE SE WITH $\ga$-DERIVATIVE}
\renewcommand{\theequation}{3.\arabic{equation}}
\setcounter{equation}{0}

Now we look at a 1-D SE with $\ga$-derivatives $d_{\ga}\sim d_{\ga}/dx$ (without motivational
physics).  We write $d_{\ga}x^{\gb}=\gb x^{\gb-\ga}$ as in (1.9) and posit
a candidate SE in the form 
\bq\label{3.1}
i\hbar\pp_t\psi=D_{\ga}\hbar^2d^2_{\ga}\psi+V(x)\psi
\end{equation}
In \cite{nott,notl} for example (cf. also \cite{cola}) one deals with a Schr\"odinger type equation
\bq\label{3.2}
{\mc D}^2\gD\psi+i{\mc D}\pp_t\psi-\frac{{\mc W}}{2m}\psi=0
\end{equation}
where ${\mc D}\sim (\hbar/2m)$ in the quantum situation.  Further ${\mc D}$ is allowed to have macro values with possible application in biology and cosmology
(see Remark 3.1 below).
\\[3mm]\indent
Consider a possible solution corresponding to $\psi=Rexp(iS/\hbar)$ in the form $({\bf 3A})\,\,\psi=RE_{\ga}\,\,(iS/\hbar)$ with $E_{\ga}$ as in (1.20).  Then one has for $S=S(x,t)\,\,({\bf 3B})\,\,\psi_t=R_t
E_{\ga}+R\pp_tE_{\ga}$ and via (1.15)-(1.16)
\bq\label{3.3}
d_{\ga}\left[RE_{\ga}\left(\frac{iS}{\hbar}\right)\right]=(d_{\ga}R)E_{\ga}+RE_{\ga}\frac{i}{\hbar}(d_{\ga}S)
\end{equation}
\bq\label{3.4}
d_{\ga}^2\left[RE_{\ga}\left(\frac{iS}{\hbar}\right)\right]=(d^2_{\ga}R)E_{\ga}+2(d_{\ga}R)E_{\ga}\frac{i}{\hbar}d_{\ga}S+
\end{equation}
$$+RE_{\ga}(\frac{i}{\hbar}d_{\ga}S)^2+RE_{\ga}
\frac{i}{\hbar}d^2_{\ga}S$$
\bq\label{3.5}
\pp_tE_{\ga}(z)=\pp_t\sum_0^{\infty}\frac{(z^{\ga}/\ga))^k}{\gG(k+1}=\frac{z_t}{\ga}\sum_1^{\infty}
\frac{(z^{\ga}/\ga)}{\gG(k)}=
\end{equation}
$$=\frac{z_t}{\ga}\sum_0^{\infty}\frac{(z^{\ga}/\ga)^m}{\gG(m+1)}=\frac{z_t}{\ga}E_{\ga}$$
Then from ({\bf 3B}), (3.4), (3.3), and (3.5) we combine real and imaginary parts in
\bq\label{3.6}
i\hbar\left[R_tE_{\ga}+\frac{iS_t}{\ga \hbar}RE_{\ga}\right]= VRE_{\ga}+
\end{equation}
$$D_{\ga}\hbar^2\left[(d_{\ga}^2R)E_{\ga}+2(d_{\ga}R)E_{\ga}\frac{i}{\hbar}d_{\ga}S-\frac{RSE_{\ga}}{\hbar^2}(d_{\ga}S)^2
+\frac{iRE_{\ga}}{\hbar}d_{\ga}^2S\right]$$
leading to
\bq\label{3.7}
R_tE_{\ga}=-2D_{\ga}d_{\ga}RE_{\ga}(d_{\ga}S)-D_{\ga}RE_{\ga}d_{\ga}^2S;
\end{equation}
$$-\frac{1}{\ga}S_tRE_{\ga}=VRE_{\ga}+D_{\ga}\hbar^2d_{\ga}^2RE_{\ga}-RE_{\ga}
(d_{\ga}S)^2$$
Thus $E_{\ga}$ cancels and we have
\bq\label{3.8}
R_t=-2D_{\ga}(d_{\ga}R)(d_{\ga}S)-D_{\ga}Rd^2_{\ga}S;
\end{equation}
$$-\frac{1}{\ga}S_tR=VR+D_{\ga}\hbar^2d_{\ga}^2R-R(d_{\ga}S)^2$$
Now recall the classical situation here as (cf. \cite{c006,c007})
\bq\label{3.9}
S_t+\frac{S_x^2}{2m}+V-\frac{\hbar^2R''}{2mR}=0;\,\,\pp_t(R^2)+\frac{1}{m}(R^2S')'=0
\end{equation}
\\[3mm]\indent
This gives an obvious comparison:
\begin{enumerate}
\item
Compare $2RR_t+(1/m)(2RR'S' +R^2 S'')=0\sim 2R_t+(1/m)(2R'S' +RS'')=0$ with 
$R_t=-2D_{\ga}(d_{\ga}R)(d_{\ga}S)-D_{\ga}Rd_{\ga}^2S$
\item
Compare $S_t+(S_x^2/2m)+V-\frac{\hbar^2R''}{2mR}=0$ with $-\frac{1}{\ga}S_t=V-
\frac{D_{\ga}\hbar^2d^2_{\ga}R}{R}+(d_{\ga}S)^2$
\end{enumerate}
which leads to 
\\[3mm]\indent
{\bf THEOREM 3.1.}  The assumption (3.1) for a 1-D $\ga$-derivative Schr\"odinger type equation
leads to a fractional quantum potential
\bq\label{3.10}
Q_{\ga}=-\frac{D_{\ga}\hbar^2d_{\ga}^2R}{R}
\end{equation}
For the classical case with $d_{\ga}R\sim R'$ (i.e. $\ga=1$) one has $D_{\ga}=1/2m$ and
one imagines more generally that $D_{\ga}\hbar^2$ may have macro values. 
 $\bs$
\\[3mm]
\indent
{\bf 	REMARK 3.1.}
We note that the techniques of scale relativity (cf. \cite{nott,notl} lead to quantum mechanics (QM).
In the non-relativistic case the fractal Hausdorff dimension $d_H=2$ arises and one can 
generate the standard quantum potential (QP) directly (cf. also \cite{cola}).  The QP turns out
to be a critical factor in understanding QM (cf. \cite{c006,c007,c009,fred,garb,gros}) while various macro
versions of QM have been suggested in biology, cosmology, etc. (cf. \cite{aunl,nott,notl,zak,zaak}).
The sign of the QP serves to distinguish diffusion from an equation with a structure forming energy term
(namely QM for $D_{\ga}=1/2m$ and fractal paths of Hausdorff dimension 2).
The multi-fractal universe of \cite{ccgn,cgot} can involve fractional calculus with various degrees
$\ga$ (i.e. fractals of differing Hausdorff dimension).
We have shown that, given a physical input for (3.1) with the $\ga$-derivative of Kobelev (\cite{kblv}), the accompanying $\ga$-QP could be related to structure formation in the
related theory.  $\bs$

\end{document}